\documentclass[aps,pre,reprint,showpacs]{revtex4-1}
\usepackage{amsmath,amssymb}
\usepackage{color}
\usepackage{graphicx}
\usepackage{hyperref}
\usepackage[utf8]{inputenc}

\usepackage{Learning}

\newcommand{\Section}[1]{\emph{#1}. --}

\begin{document}

 \title{Stochastic Thermodynamics of Learning}
 \author{Sebastian Goldt}
 \email{goldt@theo2.physik.uni-stuttgart.de} 
 \author{Udo Seifert}
 \affiliation{II. Institut für Theoretische Physik, Universität Stuttgart, 70550
   Stuttgart, Germany} 
 \date{\today}


\begin{abstract}
  Virtually every organism gathers information about its noisy environment and
  builds models from that data, mostly using neural networks. Here, we use
  stochastic thermodynamics to analyse the learning of a classification rule by
  a neural network. We show that the information acquired by the network is
  bounded by the thermodynamic cost of learning and introduce a learning
  efficiency $\eta\le1$. We discuss the conditions for optimal learning and
  analyse Hebbian learning in the thermodynamic limit.
\end{abstract}

\maketitle

\Section{Introduction} Information processing is ubiquitous in
biological systems, from single cells measuring external concentration gradients
to large neural networks performing complex motor control tasks. These systems
are surprisingly robust, despite the fact that they are operating in noisy
environments~\cite{barkai1997,bialek2011}, and they are efficient:
\emph{E. coli}, a bacterium, is near-perfect from a thermodynamic perspective in
exploiting a given energy budget to adapt to its
environment~\cite{lan2012}. Thus it is important to keep energetic
considerations in mind for the analysis of computations in living
systems. Stochastic thermodynamics~\cite{Seifert2012,parrondo2015} has emerged
as an integrated framework to study the interplay of information processing and
dissipation in interacting, fluctuating systems far from equilibrium. Encouraged
by a number of intriguing results from its application to bacterial
sensing~\cite{Qian2005,Tu2008,mehta2012,depalo2013,govern2014,
  govern2014a,barato2014a,lang2014,sartori2014,ito2015} and biomolecular
processes~\cite{andrieux2008,murugan2012,hartich2015a,lahiri2015,barato2015a},
here we consider a new problem: learning.

Learning is about extracting models from sensory data. In living systems, it is
implemented in neural networks where vast numbers of neurons communicate with
each other via action potentials, the electric pulse used universally as the
basic token of communication in neural systems~\cite{Kandel2000}. Action
potentials are transmitted via synapses, and their strength determines whether
an incoming signal will make the receiving neuron trigger an action potential of
its own. Physiologically, the adaptation of these synaptic strengths is a main
mechanism for memory formation.

\begin{figure}
  \centering
  \includegraphics[width=.8\columnwidth]{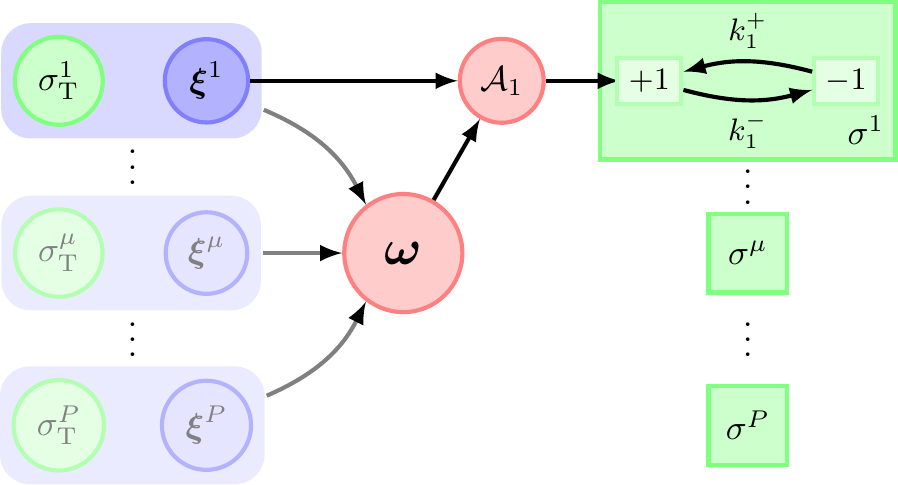}
  \caption{\label{fig:neuron}\textbf{Model of a single neuron}. Given a set of
    inputs $\vsample[\mu]\in\{\pm1\}^N$ and their true labels $\tlab[\mu]=\pm1$
    (left), the neuron learns the mappings $\vsample[\mu]\rightarrow\tlab[\mu]$
    by adjusting its weights $\vw\in\mathbb{R}^N$. It processes an input by
    computing the activation $\act^\mu~=~\vw\cdot\vsample[\mu]/\sqrt{N}$ which
    determines the transition rates of a two-state random process
    $\lab[\mu]=\pm1$ indicating the label predicted by the neuron for each
    sample, shown here for $\mu=1$.}
\end{figure}

\Section{Learning task and model} A classic example for neurons performing
associative learning are the Purkinje cells in the
cerebellum~\cite{Marr1969,Albus1971}. We model such a neuron as a single-layer
neural network or perceptron~\cite{engel2001,mackay2003}, well known from
machine learning and statistical physics~\cite{Note1}. The neuron makes $N$
connections to other neurons and is fully characterized by the weights or
synaptic strengths $\vw\in\mathbb{R}^N$ of these connections, see
figure~\ref{fig:neuron}. The neuron must learn whether it should fire an action
potential or not for a set of $P$ fixed input patterns or samples
$\vsample[\mu]=(\sample[\mu]_1,\dots,\sample[\mu]_N)$, $\mu=1,2,\dots,P$. Each
pattern describes the activity of all the other connected neurons at a point in
time: if the $n$-th connected neuron is firing an action potential in the
pattern $\vsample[\mu]$, then $\sample[\mu]_n=1$. For symmetry reasons, we set
$\sample[\mu]_n=-1$ in case the $n$-th neuron is silent in the $\mu$-th
pattern. Every sample $\vsample[\mu]$ has a fixed true label $\tlab[\mu]=\pm1$,
indicating whether an action potential should be fired in response to that input
or not. These labels are independent of each other and equiprobable; once
chosen, they remain fixed.

We model the label predicted by a neuron for each input $\vsample[\mu]$ with a
stochastic process $\lab[\mu]=\pm1$ (right panel in
figure~\ref{fig:neuron}). Assuming a thermal environment at fixed
temperature $T$, the transition rates $k_\mu^\pm$ for these processes obey the
detailed balance condition
\begin{equation}
  \label{eq:dbc}
  k_\mu^+/k_\mu^- = \exp \left(\act^\mu/\kB T\right) 
\end{equation}
where $\kB$ is Boltzmann's constant and $\act^\mu$ is the input-dependent
activation
\begin{equation}
  \label{eq:activation}
  \act^\mu\equiv\frac{1}{\sqrt{N}}\vw\cdot\vsample[\mu]
\end{equation}
where the prefactor ensures the conventional normalisation. We interpret
$p(\lab[\mu]=1\,|\,\vw)$ with fixed $\vsample[\mu]$ as the probability that the
$\mu$-th input would trigger an action potential by the neuron. The goal of
learning is to adjust the weights of the network $\vw$ such that the predicted
labels at any one time $\labs=\left(\lab[1],\dots,\lab[P]\right)$ equal the true
labels $\tlabs=\left(\tlab[1],\dots,\tlab[P]\right)$ for as many inputs as
possible.

Let us introduce the concept of learning efficiency by considering a network
with a single weight learning one sample $\sample=\pm1$ with label $\tlab$,
\emph{i.e.}  $N=P=1$. Here and throughout this letter, we set $\kB=T=1$ to
render energy and entropy dimensionless. The weight $\w(t)$ obeys an overdamped
Langevin equation~\cite{vankampen1992}
\begin{equation}
  \label{eq:langevin_n1}
  \dot{\w}(t) = - \w(t) + f(\w(t), \sample, \tlab, t) + \zeta(t).
\end{equation}
The total force on the weight arises from a harmonic potential $V(\w)=\w^2/2$,
restricting the size of the weight~\cite{Note2}, and an external force
$f(\cdot)$ introducing correlations between weight and input. The exact form of
this ``learning force'' $f(\cdot)$ depends on the learning algorithm we
choose. The thermal noise $\zeta(t)$ is Gaussian with correlations
${\avg{\zeta(t)\zeta(t')}=2\delta(t-t')}$. Here and throughout, we use angled
brackets to indicate averages over noise realisations, unless stated
otherwise. We assume that initially at $t_0=0$, the weight is in thermal
equilibrium, $p(\w)\propto\exp(-\w^2/2)$, and the labels are equiprobable,
$p(\tlab)=p(\lab)=1/2$. Choosing symmetric rates,
\begin{equation}
  \label{eq:rates}
  k^\pm=\gamma\exp(\pm\act/2),
\end{equation}
the master equation~\cite{vankampen1992} for the probability
distribution $p(\tlab, \w, \lab, t)$ with given $\sample$ reads
\begin{equation}
  \label{eq:master_n1}
  \partial_t p(\tlab, \w, \lab, t)=-\partial_\w j_\w(t) + j_{\lab}(t),
\end{equation}
where $\partial_t\equiv\partial/\partial t$ etc. and
\begin{subequations}
  \label{eq:currents_n1}
  \begin{align}
    j_\w(t)=& \left[-\w+f(\w, \sample, \tlab, t) - \partial_\w\right] p(\tlab,
              \w, \lab, t), \label{eq:currents_w_n1} \\
    j_{\lab}(t) =& k^{\lab} p(\tlab, \w, -\lab, t) - k^{-\lab} p(\tlab, \w, \lab, t) 
  \end{align}
\end{subequations}
are the probability currents for the weight and the predicted label,
respectively. In splitting the total probability current for the system
$(\tlab,\w,\lab)$ into the currents~\eqref{eq:currents_n1}, we have used the
bipartite property of the system, \emph{i.e.} that the thermal noise in each
subsystem ($\w$ and $\lab$), is independent of the
other~\cite{hartich2014,horowitz2015}.  We choose $\gamma\gg1$, \emph{i.e.}
introduce a time-scale separation between the weights and the predicted labels,
since a neuron processes a single input much faster than it
learns.

\Section{Efficiency of learning} The starting point to consider both the
information-processing capabilities of the neuron and its non-equilibrium
thermodynamics is the Shannon entropy of a random variable $X$ with probability
distribution $p(x)$,
\begin{equation}
  \label{eq:shannon}
  S(X) \equiv -\sum_{x\in X} p(x)\ln p(x),
\end{equation} 
which is a measure of the uncertainty of $X$~\cite{cover2006}. This definition
carries over to continuous random variables, where the sum is replaced by an
integral. For dependent random variables $X$ and $Y$, the conditional entropy
of $X$ given $Y$ is given by $S(X|Y)\equiv-\sum_{x,y}p(x,y)\ln p(x|y)$
where $p(x|y)=p(x,y)/p(y)$. The natural quantity to measure the information
learnt is the mutual information
\begin{equation}
  \label{eq:mutual-info}
  \mutual{\tlab}{\lab} \equiv S(\tlab) - S(\tlab|\lab)
\end{equation}
which measures by how much, on average, the uncertainty about $\tlab$ is reduced
by knowing $\lab$~\cite{cover2006}. To discuss the efficiency of learning, we
need to relate this information to the thermodynamic costs of adjusting the
weight during learning from $t_0=0$ up to a time $t$, which are given by the
well-known total entropy production~\cite{Seifert2012} of the weight,
\begin{equation}
  \label{eq:deltaStot_n1}
  \Delta S^\tot_\w \equiv \Delta S(\w) + \Delta Q.
\end{equation}
Here, $\Delta Q$ is the heat dissipated into the medium by the dynamics of the
weight and $\Delta S(\w)$ is the difference in Shannon
entropy~\eqref{eq:shannon} of the marginalized distribution
$p(\w, t)=\sum_{\tlab,\lab}p(\tlab,\w,\lab,t)$ at times $t_0$ and $t$,
respectively. 
We will show that in feedforward neural networks with Markovian
dynamics~(\ref{eq:master_n1},~\ref{eq:currents_n1}), the information learnt is
bounded by the thermodynamic costs of learning,
\begin{equation}
  \label{eq:inequality_n1}
  \mutual{\tlab}{\lab} \le \Delta S(\w) + \Delta Q
\end{equation}
for arbitrary learning algorithm $f(\w,\sample,\tlab, t)$ at all times
$t>t_0$. This inequality is our first result. We emphasise that while relations
between changes in mutual information and total entropy production have appeared
in the
literature~\cite{allahverdyan2009,sagawa2010,hartich2014,horowitz2014,horowitz2015},
they usually concern a single degree of freedom, say $X$, in contact with some
other degree(s) of freedom $Y$, and relate the change in mutual information
$\mutual{X}{Y}$ due to the dynamics of $X$ to the total entropy production of
$X$. Instead, our relation connects the entropy production in the weights with
the total change in mutual information between $\tlab$ and $\lab$, which is key
for neural networks. Our derivation~\cite{Note3} builds on recent work by
Horowitz~\cite{horowitz2015} and can be generalized to $N$ dimensions and $P$
samples, see eq.~\eqref{eq:inequality} below.  Equation \eqref{eq:inequality_n1}
suggests to introduce an efficiency of learning
\begin{equation}
  \label{eq:efficiency_n1}
  \eta \equiv  \frac{\mutual{\tlab}{\lab}}{\Delta S(\w) + \Delta Q} \le 1.
\end{equation}

\Section{Toy model} As a first example, let us calculate the efficiency of
Hebbian learning, a form of coincidence learning well known from
biology~\cite{hebb1949,Kandel2000}, for $N=P=1$ in the limit
$t\rightarrow\infty$. If the neuron should fire an action potential when its
input neuron fires, or if they should both stay silent, \emph{i.e.}
$\sample=\tlab=\pm1$, the weight of their connection increases -- ``fire
together, wire together''. For symmetry reasons, the weight decreases if the
input neuron is silent but the neuron should fire and vice versa,
$\sample=-\tlab$. This rule yields a final weight proportional to
$\mathcal{F}\equiv\tlab \sample$, so to minimise dissipation~\cite{abreu2011},
we choose a learning force $f$ linearly increasing with time,
\begin{equation}
  \label{eq:hebbian-force}
  f(\w,\sample,\tlab, t) \equiv
  \begin{cases}
    \lr \mathcal{F} t / \tau & t \le \tau \\
    \lr \mathcal{F} & t > \tau,
  \end{cases}
\end{equation}
where we have introduced the learning duration $\tau>0$ and the factor $\lr>0$
is conventionally referred to as the learning rate in the machine learning
literature~\cite{engel2001}. The total entropy
production~\eqref{eq:deltaStot_n1} can be computed from the distribution
$p(\tlab, \w, t)$, which is obtained by first integrating $\lab$ out of
equations~(\ref{eq:master_n1},~\ref{eq:currents_n1}) and solving the resulting
Fokker-Planck equation~\cite{risken1996}. The total heat dissipated into the
medium $\Delta Q$ is given by~\cite{Seifert2012}
\begin{multline}
  \label{eq:deltaQ}
  \Delta Q = \int_0^\infty\diff{t} \int_{-\infty}^\infty  \diff{\w} \;
  j_\w(t) \left[- \w(t) + f(\w(t), \sample, \tlab, t)\right]\\ = \frac{\lr^2\mathcal{F}^2(e^{-\tau}+\tau-1)}{\tau^2}.
\end{multline}
As expected, no heat is dissipated in the limit of infinitely slow driving,
$\lim_{\tau\rightarrow\infty}\Delta Q = 0$, while for a sudden potential
switch $\tau\rightarrow0$,
$\lim_{\tau\rightarrow0}\Delta Q = \lr^2\mathcal{F}^2/2$. The change in
Shannon entropy $\Delta S(\w)$ is computed from the marginalized distribution
$p(\w, t)=\sum_{\tlab} p(\tlab, \w, t)$.  Finally, the mutual
information~\eqref{eq:mutual-info} can be computed from the stationary solution
of~\eqref{eq:master_n1}.

\begin{figure}
  \centering 
  \includegraphics[width=.66\columnwidth]{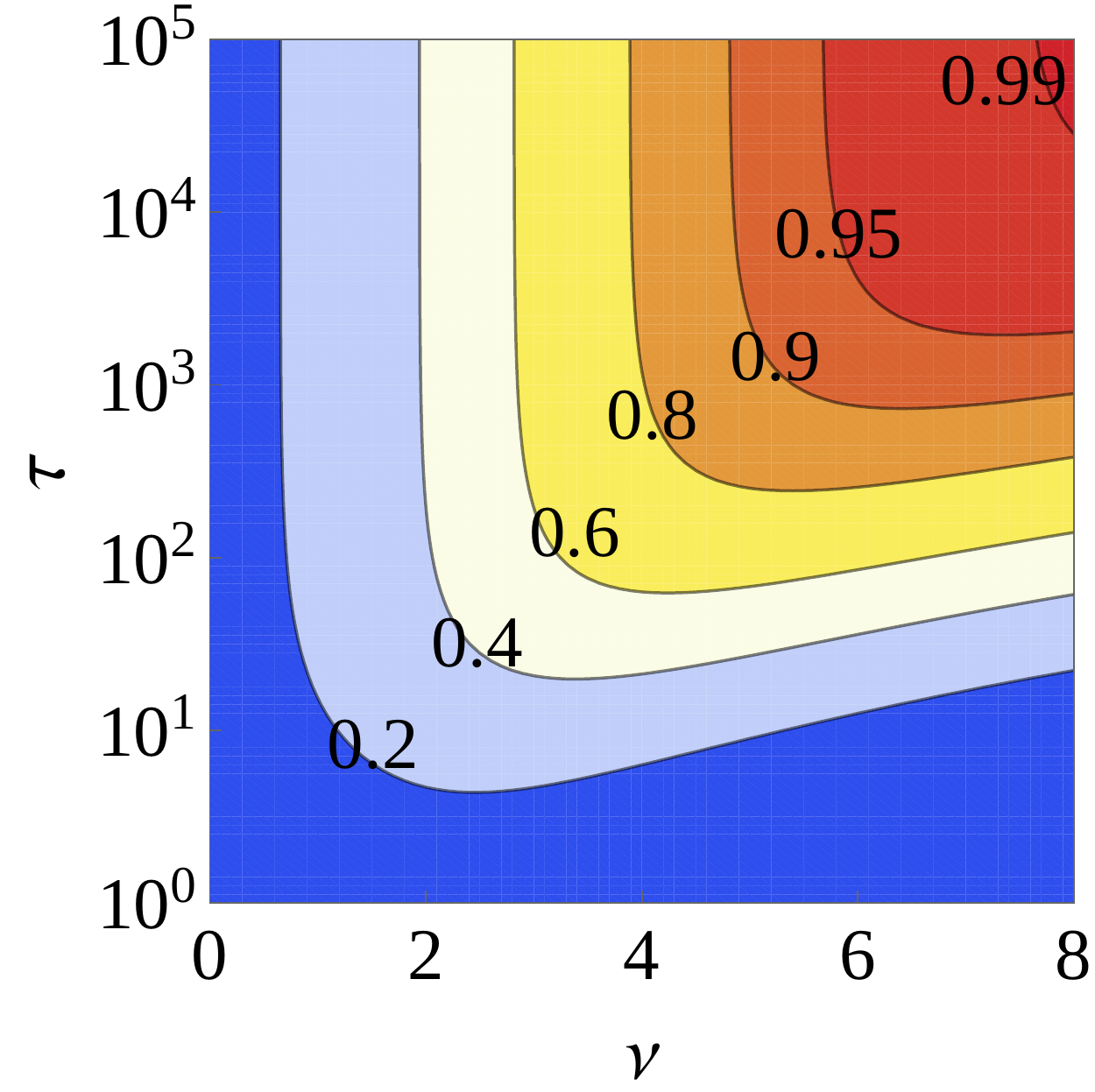}
  \caption{\label{fig:efficiency_n1}\textbf{Learning efficiency of a neuron with
      a single weight.} We plot the efficiency $\eta$ ~\eqref{eq:efficiency_n1}
    for a neuron with a single weight learning a single sample as a function of
    the learning rate $\lr$ and learning duration $\tau$ in the limit
    $t\rightarrow\infty$.}
\end{figure}

A plot of the efficiency~\eqref{eq:efficiency_n1}, fig.~\ref{fig:efficiency_n1},
highlights the two competing requirements for maximizing $\eta$. First, all the
information from the true label $S(\tlab)=\ln 2$ needs to be stored in the
weight by increasing the learning rate $\lr$, which leads to $\Delta S(\w)\rightarrow\ln 2$ and a
strongly biased distribution $p(\lab|\w)$ such that
$\mutual{\tlab}{\lab}\rightarrow\ln 2$. Second, we need to minimise the
dissipated heat $\Delta Q$, which increases with $\lr$, by driving the weight
slowly, $\tau\gg1$.

\Section{More samples, higher dimensions} Moving on to a neuron with $N$ weights
$\vw$ learning $P$ samples with true labels
$\tlabs\equiv(\tlab[1],\dots,\tlab[\mu],\dots,\tlab[P])$, we have a Langevin equation for each
weight $\wn$ with independent thermal noise sources $\zeta_n(t)$ such that
$\avg{\zeta_n(t)\zeta_m(t')}=2\delta_{nm}\delta(t-t')$ for $n,m=1,\dots,N$. Two
learning scenarios are possible: \emph{batch learning}, where the learning force
is a function of all samples and their labels,
\begin{equation}
  \label{eq:langevin_batch}
  \dot{\w}_n(t) = - \wn(t) + f(\wn(t), \{\sample[\mu]_n, \tlab[\mu]\}, t) + \zeta_n(t).
\end{equation}
A more realistic scenario from a biological perspective is \emph{online
  learning}, where the learning force is a function of only one sample and its
label at a time,
\begin{equation}
  \label{eq:langevin_online}
  \dot{\w}_n(t) = - \wn(t) +  f(\wn(t), \sample[\mu(t)]_n, \tlab[\mu(t)], t) + \zeta_n(t).
\end{equation}
The sample and label which enter this force are given by
$\mu(t)\in\{1,\dots,P\}$, which might be a deterministic function or a random
process. Either way, the weights $\vw$ determine the transition rates of the $P$
independent two-state processes for the predicted labels
$\labs\equiv(\lab[1],\dots,\lab[\mu],\dots,\lab[P])$ via~\eqref{eq:dbc}
and~\eqref{eq:activation}. Again, we assume that the thermal noise in each
subsystem, $\wn$ or $\lab[\mu]$, is independent of all the others, and choose
initial conditions at $t_0=0$ to be $p(\vw)\propto\exp(-\vw\cdot\vw/2)$ and
$p(\tlab[\mu])=p(\lab[\mu])=1/2$. The natural quantity to measure the amount of
learning after a time $t$ in both scenarios is the sum of
$\mutual{\tlab[\mu]}{\lab[\mu]}$ over all inputs. We can show~\cite{Note3} that
this information is bounded by the total entropy production of all the weights,
\begin{equation}
  \label{eq:inequality}
  \sum_{\mu=1}^P \mutual{\tlab[\mu]}{\lab[\mu]} \le  \sum_{n=1}^N \left[ \Delta S(\wn) +
    \Delta Q_n\right] = \sum_{n=1}^N \Delta S_n^\textrm{tot}
\end{equation}
where $\Delta Q_n$ is the heat dissipated into the medium by the $n$-th weight
and $\Delta S(\wn)$ is the change from $t_0$ to $t$ in Shannon
entropy~\eqref{eq:shannon} of the marginalized distribution $p(\wn, t)$. This is
our main result.

Let us now compute the efficiency of online Hebbian learning in the limit
$t\rightarrow\infty$. Since a typical neuron will connect to $\sim1000$ other
neurons~\cite{Kandel2000}, we take the thermodynamic limit by letting the number
of samples $P$ and the number of dimensions $N$ both go to infinity while
simultaneously keeping the ratio
\begin{equation}
  \label{eq:alpha}
  \alpha\equiv P/N  
\end{equation}
on the order of one. The samples $\vsample[\mu]$ are drawn at random from
$p(\sample[\mu]_n=1)=p(\sample[\mu]_n=-1)=1/2$ and remain fixed~\cite{Note4}.
We choose a learning force on the $n$-th weight of the form
\eqref{eq:hebbian-force} with $\mathcal{F}\rightarrow\mathcal{F}_n$ and assume
that the process $\mu(t)$ is a random walk over the integers $1,\dots,P$
changing on a timescale much shorter than the relaxation time of the
weights. Since $f^2$ is finite, the learning force is effectively constant with
\begin{equation}
  \label{eq:final_weight_mean}
  \mathcal{F}_n = \frac{1}{\sqrt{N}} \sum_{\mu=1}^P \sample[\mu]_n \tlab[\mu],
\end{equation} 
where the prefactor ensures the conventional
normalisation~\cite{engel2001}. Hence all the weights $\wn$ are independent of
each other and statistically equivalent. Averaging first over the noise with
fixed $\tlabs$, we find that $\wn$ is normally distributed with mean
$\avg{\wn}=\lr\mathcal{F}_n$ and variance 1~\cite{Note5}. The average with
respect to the quenched disorder $\tlabs$, which we shall indicate by an
overline, is taken second by noting that $\mathcal{F}_n$ is normally distributed
by the central limit theorem with $\overline{\mathcal{F}_n}=0$ and
$\overline{\mathcal{F}^2_n}=\alpha$, hence $\overline{\avg{\wn}}=0$ and
$\overline{\avg{\w_n^2}}=1+\alpha\lr^2$. The change in Shannon entropy of the
marginalized distribution $p(\wn)$ is hence $\Delta S(\wn)=\ln(1+\alpha\lr^2)$.
Likewise, the heat dissipated by the $n$-th weight $\overline{\Delta Q_n}$ is
obtained by averaging eq.~\eqref{eq:deltaQ} over
$\mathcal{F}\rightarrow\mathcal{F}_n$.

\begin{figure}
  \centering
  \includegraphics[width=.78\columnwidth]{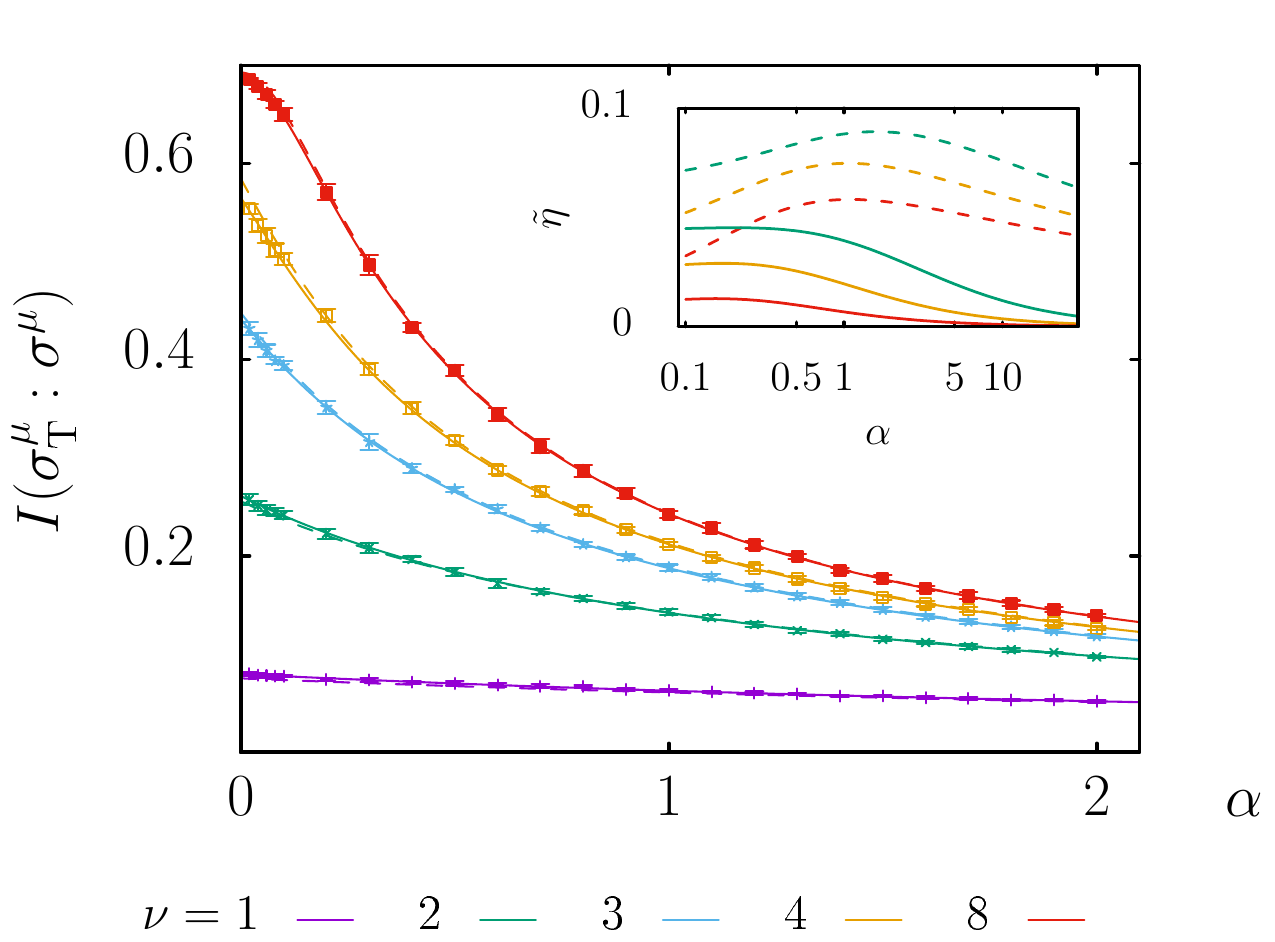}
  \caption{\label{fig:efficiency_thermo}\textbf{Hebbian learning in the
      thermodynamic limit.} We plot the mutual information between the true and
    predicted label of a randomly chosen sample \eqref{eq:mutual_thermo} in the
    limit $t\rightarrow\infty$ with $N,P\rightarrow\infty$ as a function of
    $\alpha\equiv P/N$, computing $\pC^\mu$ from~\eqref{eq:pR} (solid lines) and
    by Monte Carlo integration of $p(\tlabs,\vw,\labs)$ (crosses, error bars
    indicate one standard deviation). The inset shows the learning
    efficiency~\eqref{eq:efficiency_typical} in the limits $\tau\rightarrow0$
    (solid) and $\tau\rightarrow\infty$ (dashed). In both plots, $\lr$ increases
    from bottom to top.}
\end{figure}

The mutual information $\mutual{\tlab[\mu]}{\lab[\mu]}$ is a functional of the
marginalized distribution $p(\tlab[\mu], \lab[\mu])$ which can be obtained by
direct integration of $p(\tlabs, \vw, \labs)$~\cite{Note3}. Here we will take a
simpler route starting from the \emph{stability} of the $\mu$-th
sample~\cite{gardner1987}
\begin{equation}
  \label{eq:stability}
  \Delta^\mu \equiv \frac{1}{\sqrt{N}} \vw\cdot\vsample[\mu]\tlab[\mu] = \act^\mu \tlab[\mu].
\end{equation}
Its role can be appreciated by considering the limit $T\rightarrow0$, where it
is easily verified using the detailed balance condition~\eqref{eq:dbc} that the
neuron predicts the correct label \iff $\Delta^\mu>0$. For $T=1$, the neuron
predicts the $\mu$-th label correctly with probability
\begin{equation}
  \label{eq:pR}
  \pC^\mu \equiv p(\lab[\mu]=\tlab[\mu]) = \int_{-\infty}^\infty \dd \! \Delta^\mu \; p(\Delta^\mu)\frac{e^{\Delta^\mu}}{e^{\Delta^\mu}+1}
\end{equation}
where $p(\Delta^\mu)$ is the distribution generated by thermal noise and
quenched disorder, yielding a Gaussian with mean $\lr$ and
variance~$1+\alpha\lr^2$~\cite{Note3}. The mutual information follows as
\begin{equation}
  \label{eq:mutual_thermo}
  \mutual{\tlab[\mu]}{\lab[\mu]} = \ln 2 - S(\pC^\mu)
\end{equation}
with the shorthand for the entropy of a binary random variable
$S(p)=-p \ln p-(1-p)\ln(1-p)$~\cite{cover2006}. It is plotted in
fig.~\ref{fig:efficiency_thermo} together with the mutual information obtained
by Monte Carlo integration of $p(\tlabs,\vw,\labs)$ with $N=10 000$. For a
vanishing learning rate $\lr\rightarrow0$ or infinitely many samples
$\alpha\rightarrow\infty$, $\pC^\mu \rightarrow1/2$ and hence
$\mutual{\tlab[\mu]}{\lab[\mu]}\rightarrow0$. The maximum value
$\mutual{\tlab[\mu]}{\lab[\mu]}=\ln 2$ is only reached for small $\alpha$ and
decreases rapidly with increasing $\alpha$, even for values of $\alpha$ where it
is possible to construct a weight vector that classifies all the samples
correctly~\cite{mackay2003}. This is a consequence of both the thermal noise in
the system and the well-known failure of Hebbian learning to use the information
in the samples perfectly~\cite{engel2001}. We note that while the integral in
eq.~\eqref{eq:pR} has to be evaluated numerically, $\pC^\mu$ can be closely
approximated analytically by $p(\Delta^\mu>0)$ with the replacement
$\lr\rightarrow\lr/2$~\cite{Note3} (dashed lines in
fig.~\ref{fig:efficiency_thermo}).

Together, these results allow us to define the efficiency $\tilde{\eta}$ of
Hebbian learning as a function of just $\alpha$ and $\lr$,
\begin{equation}
  \label{eq:efficiency_typical}
  \tilde{\eta} \equiv \alpha \frac{\mutual{\tlab[\mu]}{\lab[\mu]}}{\Delta S(\wn) + \overline{\Delta Q_n}},
\end{equation}
where we have taken the mutual information per sample and the total entropy
production per weight, multiplied by the number of samples and weights,
respectively. Plotted in the inset of figure~\ref{fig:efficiency_thermo}, this
efficiency never reaches the optimal value 1, even in the limit of vanishing
dissipation $\tau\rightarrow\infty$ (solid lines in
fig. \ref{fig:efficiency_thermo}).

\Section{Conclusion and perspectives} We have introduced neural networks as
models for studying the thermodynamic efficiency of learning. For the
paradigmatic case of learning arbitrary binary labels for given inputs, we
showed that the information acquired is bounded by the thermodynamic cost of
learning. This is true for learning an arbitrary number of samples in an
arbitrary number of dimensions for any learning algorithm without feedback for
both batch and online learning.

Our framework opens up numerous avenues for further work. It will be interesting
to analyse the efficiency of learning algorithms that employ feedback or use an
auxiliary memory~\cite{hartich2016}. Furthermore, synaptic weight distributions
are experimentally accessible~\cite{brunel2004,barbour2007}, offering the
exciting possibility to test predictions on learning algorithms by looking at
neural weight distributions. The inverse problem, \emph{i.e.}  deducing features
of learning algorithms or the neural hardware that implements them by optimising
some functional like the efficiency, looks like a formidable challenge, despite
some encouraging progress in related fields~\cite{tkacik2009,sokolowski2015}.

\begin{acknowledgments}
  We thank David Hartich for stimulating discussions and careful reading of the
  manuscript.
\end{acknowledgments}

\bibliographystyle{aip}

\pagebreak
\onecolumngrid

\setcounter{equation}{0}
\setcounter{figure}{0}
\setcounter{table}{0}
\setcounter{page}{1}
\makeatletter
\renewcommand{\theequation}{S\arabic{equation}}
\renewcommand{\thefigure}{S\arabic{figure}}
\renewcommand{\bibnumfmt}[1]{[S#1]}
\renewcommand{\citenumfont}[1]{S#1}

\section*{\large \textbf{Supplemental Material}\\Stochastic Thermodynamics of
  Learning}

\begin{center}
  Sebastian Goldt and Udo Seifert\\
  \emph{II. Institut für Theoretische Physik, Universität Stuttgart, 70550
    Stuttgart, Germany}\\
  (Dated: \today)
\end{center}

  In this supplemental material, we discuss the stochastic thermodynamics of
  neural networks in detail in section~\ref{sec:stochastic-thermodyanmics} and
  derive our main result, eq.~\eqref{eq:inequality} of the main text, in
  section~\ref{sec:inequalities}. Furthermore, we complement our discussion Hebbian
  learning in the thermodynamic limit with additional analytical calculations in
  section~\ref{sec:hebbian-calculations}.

\section{Stochastic thermodynamics of neural networks}
\label{sec:stochastic-thermodyanmics}

We now give a detailed account of the stochastic thermodynamics of neural
networks. For simplicity, here we will focus on batch learning; the
generalisation to online learning is straightforward. For a network with $N$
weights $\wn\in\mathbb{R}^N$ learning $P$ samples $\vsample[\mu]\in\{\pm1\}^N$
with their labels $\tlab[\mu]=\pm1$, $\mu=1,2,\dots,P$, we have $N$ Langevin
equations~\cite{s_vankampen1992}
\begin{equation}
  \label{eq:langevin}
  \dot{\w}_n(t) = - \wn(t) + f(\wn(t), \{\sample[\mu]_n, \tlab[\mu]\}, t) + \zeta_n(t).
\end{equation}
The Gaussian noise $\zeta_n(t)$ has correlations
$\avg{\zeta_n(t)\zeta_m(t')}=2T\delta_{nm}\delta(t-t')$ for $n,m=1,\dots,N$
where $T$ is the temperature of the surrounding medium and we have set
Boltzmann's constant to unity to render entropy dimensionless. The
weights $\vw$ determine the transition rates of the $P$ independent two-state
processes for the predicted labels $\lab[\mu]$ via
\begin{equation}
  k_\mu^+/k_\mu^- = \exp \left(\act^\mu/T\right) 
\end{equation}
where $\act^\mu$ is the input-dependent activation
\begin{equation}
  \act^\mu\equiv\frac{1}{\sqrt{N}}\vw\cdot\vsample[\mu]
\end{equation}
For the remainder of this supplemental material, we set $T=1$, rendering energy
dimensionless. We assume that the thermal noise in each subsystem, like $\wn$ or
$\lab[\mu]$, is independent of all the others. This multipartite
assumption~\cite{s_horowitz2015} allows us to write the master equation for the
distribution $p(\tlabs,\vw,\labs, t)$ with
$\tlabs\equiv(\tlab[1],\dots,\tlab[P])$ and $\labs\equiv(\lab[1],\dots,\lab[P])$
as
\begin{equation}
  \label{eq:master}
  \partial_t p(\tlabs,\vw,\labs, t)=-\sum_{n=1}^N \partial_n j_n(t) + \sum_{\mu=1}^Pj_\mu(t),
\end{equation}
where $\partial_t\equiv\partial/\partial t$,
$\partial_n\equiv \partial/\partial\, \wn$ and the probability currents for the
$n$-th weight $\wn$ and the $\mu$-th predicted label $\lab[\mu]$ are given by
\begin{subequations}
  \label{eq:currents}
  \begin{align}
    j_n(t)=& \left[-\wn+f(\wn, \vsample[\mu(t)], \tlab[\mu(t)], t) - \partial_n\right]
             p(\tlabs,\vw,\labs, t), \\
      j_\mu(t) =& k^+ p(\tlabs,\vw,\lab[1],\dots,-\lab[\mu],\dots,\lab[P], t)
      - k^- p(\tlabs, \vw, \labs, t).
  \end{align}
\end{subequations}
We choose symmetric rates $k^\pm_\mu=\gamma\exp(\pm\act^\mu/2)$ with
$\gamma\gg1$. Initially, the true labels $\tlabs$, weights $\vw$ and predicted
labels are all uncorrelated with
\begin{align}
  p_0(\tlab[\mu])=&1/2, \\
  p_0(\lab[\mu])=&1/2, \quad \text{and} \\
  p_0(\vw) =& \frac{1}{(2\pi)^{N/2}} \exp(-\vw\cdot\vw/2).
\end{align}
Since the following discussion applies to the time-dependent
dynamics~\eqref{eq:master}, we understand that all quantities that will be
introduced in the remainder of this section have an implicit
time-dependence via the distribution $p(\tlabs, \vw, \labs, t)$ or the
currents~\eqref{eq:currents}.

Our starting point for the stochastic thermodynamics of this system is the
well-known total entropy production $\dot{S}^\tot$ of the network which obeys
the following second-law like inequality~\cite{s_Seifert2012}
\begin{equation}
  \dot{S}^\tot = \partial_t S(\tlabs, \vw, \labs) + \dot{S}^\m \ge 0
\end{equation}
with equality in equilibrium only. Here, we have the Shannon
entropy~\cite{s_cover2006} of the system,
\begin{equation}
  \label{eq:shannon_all}
  S(\tlabs, \vw, \labs) = - \sum_{\tlabs,\labs} \int_{-\infty}^\infty \diff{\vw} \;
  p(\tlabs, \vw, \labs) \ln p(\tlabs, \vw, \labs).
\end{equation}
Here, we include the variables $\tlabs$, $\vw$ and $\labs$ as arguments of the
function $S$ in a slight abuse of notation to emphasise that we consider the
Shannon entropy of the full distribution $p(\tlabs, \vw, \labs)$.  $\dot{S}^\m$
gives the rate of entropy production in the medium. For a system at constant
temperature $T=1$, $\dot{S}^\m\equiv\dot{Q}$, the rate of heat dissipation into
the medium~\cite{s_Seifert2012}. Let us first focus on the change in Shannon
entropy by differentiating~\eqref{eq:shannon_all} with respect to time,
\begin{equation}
  \partial_t S(\tlabs, \vw, \labs) = - \sum_{\tlabs,\labs} \int_{-\infty}^\infty \diff{\vw} \;
  \dot{p}(\tlabs, \vw, \labs) \ln p(\tlabs, \vw, \labs),
\end{equation}
where we have used that $p(\tlabs, \vw, \labs)$ is, of course, normalised. Using
the master equation~\eqref{eq:master}, we find that
\begin{equation}
  \partial_t S(\tlabs, \vw, \labs) = \sum_{n=1}^N \dot{S}_n + \sum_{\mu=1}^P \dot{S}_\mu
\end{equation}
where 
\begin{align}
  \dot{S}_n \equiv&  \sum_{\tlabs,\labs} \int_{-\infty}^\infty \diff{\vw} \;
  \partial_n j_n(t) \ln p(\tlabs, \vw, \labs),\\
  \dot{S}_\mu \equiv& - \sum_{\tlabs,\labs} \int_{-\infty}^\infty \diff{\vw} \;
  j_\mu \ln p(\tlabs, \vw, \labs),
\end{align}
are the rate of change of the Shannon entropy $S(\tlabs, \vw, \labs)$ due to the
dynamics of $\w_n$ and $\lab[\mu]$, respectively. The key point here is that
multipartite dynamics, a consequence of the uncorrelated noise across
subsystems, lead to a linear splitting of the probability currents and hence to
a linear splitting of all quantities which are functions of the total
probability current. Similarly, for the rate of heat dissipation $\dot{Q}$,
we can write
\begin{equation}
  \dot{Q} = \sum_{n=1}^N \dot{Q}_n + \sum_{\mu=1}^P\dot{Q}_\mu
\end{equation}
where
\begin{equation}
  \dot{Q}_n =  \sum_{\tlabs,\labs} \int_{-\infty}^\infty \diff{\vw} \;
  j_n(t) F_n(\tlabs, \vw, \labs)
\end{equation}
with the total force on the $n$-th weight
$F_n=- \wn(t) + f(\wn(t), \{\sample[\mu]_n, \tlab[\mu]\}, t)$, while
\begin{equation}
\dot{Q}_\mu = \sum_{\tlabs,\labs} \int_{-\infty}^\infty \diff{\vw} \;
  j_\mu(t) \lab[\mu] \vw\cdot\vsample[\mu]/2.
\end{equation} 
Finally, total entropy production $\dot{S}^\tot$ can also be split,
\begin{equation}
  \dot{S}^\tot = \sum_{n=1}^N \dot{S}^\tot_n + \sum_{\mu=1}^P \dot{S}^\tot_\mu.
\end{equation}
It can easily be shown that each of these total entropy productions of a
subsystem obeys a separate second-law like inequality, \emph{e.g.}
\begin{equation}
  \label{eq:2nd-law-short-n}
  \dot{S}^\tot_n =\dot{S}_n(\tlabs, \vw, \labs) + \dot{Q}_n \geq 0
\end{equation}
for the $n$-th weight. 

Writing
\begin{equation}
p(\tlabs, \vw, \labs) = p(\wn)p(\tlabs, \vwOthers, \labs |\wn)
\end{equation}
with $\vwOthers\equiv(\cdots, \w_{n-1}, \w_{n+1},\cdots)$, we can split
$\dot{S}_n(\tlabs,\vw,\labs)$ into two parts: first, the change of Shannon
entropy of the marginalized distribution $p(\wn)$,
\begin{equation}
  \dot{S}_n(\wn) = \sum_{\tlabs,\labs} \int_{-\infty}^\infty \diff{\vw} \;
   \partial_n j_n(t) \ln p(\wn) = \partial_t S(\wn),
\end{equation}
where the last equality follows from the fact that an entropy change of the
marginalized distribution $p(\wn)$ can only come from the dynamics of $\wn$. The
second part is called the learning rate \cite{s_hartich2014}
\begin{equation}
  \label{eq:lw}
  l_n(\w_n; \tlabs, \labs, \vwOthers) = 
    - \sum_{\tlabs,\labs} \int_{-\infty}^\infty \diff{\vw} \; \partial_n j_n(t) \ln p(\tlabs,\labs, \vwOthers |\w_n)
\end{equation}
or information flow \cite{s_allahverdyan2009,s_horowitz2014}. We emphasise that this
learning rate $l_n$ is thermodynamic and has nothing to do with the learning
rate $\lr$ that goes into the definition of the learning algorithms, see for
example eq.~\eqref{eq:hebbian-force} of the main text. To avoid confusion, we
will refer to $l_n$ as the thermodynamic learning rate for the remainder of this
supplemental material. The second law \eqref{eq:2nd-law-short-n} for the $n$-th
weight hence becomes
\begin{equation}
  \label{eq:2nd-law-long-n}
  \dot{S}^\tot_n = \partial_t S(\wn) + \dot{Q}_n - l_n(\w_n; \tlabs, \labs, \vwOthers)\geq 0
\end{equation}
The thermodynamic learning rate is a thermodynamically consistent measure of how
much the dynamics of $\wn$ change the mutual information
$\mutual{\wn}{\tlabs, \vwOthers, \labs}$, in particular for a system that
continuously rewrites a single memory~\cite{s_horowitz2014a}.

We can further refine the second law~\eqref{eq:2nd-law-long-n} by exploiting the
causal structure of the dynamics, as was recently suggested by
Horowitz~\cite{s_horowitz2015}. The subsystem $\wn$ directly interacts only with
those degrees of freedom that appear in its probability current $j_n(t)$
\eqref{eq:currents}. From inspection of the current $j_n(t)$, we see that $\wn$
is directly influenced only by itself and the given labels $\tlabs$. Keeping
this in mind, we use the chain rule for mutual information~\cite{s_cover2006} to
write
\begin{equation}
  \mutual{\wn}{\tlabs, \vwOthers, \labs}= \mutual{\wn}{\tlabs} +
  \mutualGiven{\wn}{\vwOthers, \labs}{\tlabs},
\end{equation}
where we use the conditional mutual information
\begin{align}
  \label{eq:mutual-info-conditional}
  \mutualGiven{\wn}{\vwOthers, \labs}{\tlabs} =& S(\wn|\tlabs) -
                                                 S(\wn|\vwOthers, \labs, \tlabs) \\
 =&-
  \sum_{\labs,\tlabs}\int_{-\infty}^\infty \diff{\vw} \; p(\tlabs, \vw, \labs) \ln
  \frac{p(\tlabs, \vw, \labs)p(\tlabs)}{p(\wn, \tlabs)p(\vwOthers, \labs, \tlabs)}.
\end{align}
Accordingly, we split the thermodynamic learning rate~\eqref{eq:lw} into a
thermodynamic learning rate of the $n$-th weight with the degrees of freedom
that it directly interacts with, \emph{i.e.} the true labels $\tlabs$,
\begin{equation}
  \label{eq:lw-refined}
  l_n(\wn; \tlabs) = - \sum_{\labs,\tlabs}\int_{-\infty}^\infty \diff{\vw}
  \; \partial_n j_n(t)\ln p(\tlabs|\wn),
\end{equation}
and a thermodynamic learning rate with the other subsystems given the true labels,
\begin{equation}
  \label{eq:lw-conditional}
  l_n(\wn; \vwOthers, \labs|\tlabs) = - \sum_{\labs,\tlabs}\int_{-\infty}^\infty
    \diff{\vw} \; \partial_n j_n(t)\ln \left(\frac{p(\wn, \vwOthers, \labs
        |\tlabs)}{p(\wn|\tlabs)p(\vwOthers, \labs|\tlabs)}\right).
\end{equation}
Horowitz proved \cite{s_horowitz2015} the following second-law like inequality
including the refined thermodynamic learning rate~\eqref{eq:lw-refined},
\begin{equation}
  \label{eq:2nd-law-refined}
  \partial_t S(\wn) + \dot{Q}_n - l_n(\wn; \tlabs) \geq 0.
\end{equation}
which is the basis for our proof of the main inequality,
equation~\eqref{eq:inequality} of the main text.

\section{Derivation of inequality~$\eqref{eq:inequality}$ of the main text}
\label{sec:inequalities}

The stochastic thermodynamics of neural networks yields $N$ inequalities of the
form~\eqref{eq:2nd-law-refined}. Integrating over time and summing over all the
weights, we find
\begin{equation}
  \sum_{n=1}^N \left[ \Delta S(\wn) + \Delta Q_n\right] \ge \sum_{n=1}^N \int_0^\infty \diff{t} \; l_n(\wn; \tlabs) = \sum_{n=1}^N \Delta
  \mutual{\wn}{\tlabs}
\end{equation}
The precise definition of all the terms are discussed in the main text and in
section \ref{sec:stochastic-thermodyanmics} of this supplemental material. The
crucial point for the last equality is that the labels $\tlabs$ are static, so
that the mutual information $\mutual{\wn}{\tlabs}$ changes only due to the
dynamics of $\wn$ and hence $\partial_t \mutual{\wn}{\tlabs}=l_n(\wn;\tlabs)$
\cite{Note6}.  To make progress towards our main result,
inequality~\eqref{eq:inequality} of the main text, we need to show that
\begin{equation}
  \label{eq:1}
  \sum_{n=1}^N \Delta \mutual{\wn}{\tlabs} \ge \sum_{\mu=1}^P \Delta \mutual{\tlab[\mu]}{\lab[\mu]}.
\end{equation}

First, we note that from the chain rule of mutual information \cite{s_cover2006},
we have
\begin{equation}
  \Delta \mutual{\vw}{\tlabs}=\Delta \mutual{\w_1, \dots, \w_n}{\tlabs}=\sum_{n=1}^N\Delta \mutualGiven{\wn}{\tlabs}{\w_{n-1},\dots,\w_1}
\end{equation}
with the conditional mutual information \cite{s_cover2006}
\begin{equation}
  \mutualGiven{\wn}{\tlabs}{\w_{n-1},\dots,\w_1} \equiv S(\wn|\w_{n-1},\dots,\w_1) -
  S(\wn|\tlabs, \w_{n-1},\dots,\w_1).
\end{equation}
Due to the form of the Langevin equation for the single weight,
eq.~\eqref{eq:langevin}, individual weights are uncorrelated, and hence the
conditional mutual information simplifies to
\begin{align}
  \Delta \mutualGiven{\wn}{\tlabs}{\w_{n-1},\dots,\w_1} &=
                                                    \Delta S(\wn|\w_{n-1},\dots,\w_1)-\Delta S(\wn|\tlabs,\w_{n-1},\dots,\w_1)\\
  &=  \Delta S(\wn) - \Delta S(\wn|\tlabs)\\
  &= \Delta \mutual{\wn}{\tlabs}
\end{align}
such that
\begin{equation}
  \sum_{n=1}^N\Delta \mutual{\wn}{\tlabs} = \Delta \mutual{\vw}{\tlabs}.
\end{equation}

Next, we show that
\begin{equation}
  \label{eq:first-inequality}
  \Delta \mutual{\vw}{\tlabs} = \sum_{\mu=1}^P
  \Delta \mutualGiven{\vw}{\tlab[\mu]}{\tlab[\mu-1],\dots,\tlab[1]}\stackrel{!}{\ge} \sum_{\mu=1}^P\Delta \mutual{\vw}{\tlab[\mu]}.
\end{equation}
using the independence of the given labels $\tlabs$.  We first note that
\begin{align}
  \Delta \mutualGiven{\vw}{\tlab[\mu]}{\tlab[\mu-1],\dots,\tlab[1]}= & \Delta S(\tlab[\mu]|\tlab[\mu-1],\dots,\tlab[1])
                                                               - \Delta S(\tlab[\mu]|\vw, \tlab[\mu-1],\dots,\tlab[1]) \\
  =& \Delta S(\tlab[\mu]) - \Delta S(\tlab[\mu]|\vw, \tlab[\mu-1],\dots,\tlab[1]) 
\end{align}
while
\begin{equation}
  \Delta \mutual{\vw}{\tlab[\mu]}=\Delta S(\tlab[\mu]) - \Delta S(\tlab[\mu]|\vw)
\end{equation}
Hence for
$\Delta
\mutualGiven{\vw}{\tlab[\mu]}{\tlab[\mu-1],\dots,\tlab[1]}\stackrel{!}{\ge}\Delta
\mutual{\vw}{\tlab[\mu]}$, we need
\begin{align}
  \Delta \mutualGiven{\vw}{\tlab[\mu]}{\tlab[\mu-1],\dots,\tlab[1]}-
           \Delta \mutual{\vw}{\tlab[\mu]} & \\ 
  = \quad & \Delta S(\tlab[\mu]|\vw) - \Delta S(\tlab[\mu]|\vw,
            \tlab[\mu-1],\dots,\tlab[1]) \label{eq:first-step} \\
  = \quad & \Delta \mutualGiven{\tlab[\mu]}{\tlab[\mu-1],\dots,\tlab[1]}{\vw} \\
  \ge \quad &  0
\end{align}
where we first used that the $\tlab[\mu]$ are independent and identically
distributed. The last inequality follows since any mutual information,
conditional or not, is always greater than or equal to zero \cite{s_cover2006}. We
have thus shown that
$\Delta \mutualGiven{\vw}{\tlab[\mu]}{\tlab[\mu-1],\dots,\tlab[1]}\ge\Delta
\mutual{\vw}{\tlab[\mu]}$ and hence~\eqref{eq:first-inequality} is true.

Finally, to prove that
$\Delta \mutual{\vw}{\tlab[\mu]}>\Delta \mutual{\tlab[\mu]}{\lab[\mu]}$, we
consider the full probability distribution $p(\tlabs,\vw,\labs)$. From the
master equation, eq.~\eqref{eq:master}, we can write this distribution as
\begin{equation}
  p(\tlabs, \vw, \labs) = p(\tlabs)p(\vw|\tlabs)\left[p^{(0)}(\labs|\vw) +
    \frac{1}{\gamma}p^{(1)}(\labs|\vw) + \mathcal{O}(1/\gamma^2)\right]
\end{equation}
with $\gamma\gg1$ for physiological reasons as described in the text -- it takes
the neuron longer to learn than to generate an action potential. Hence to first
order, $\tlabs\rightarrow\vw\rightarrow\labs$ is by definition a Markov chain~\cite{s_cover2006}. Integrating out all the labels, true and predicted, except
for the $\mu$-th one, we have the Markov chain
$\tlab[\mu]\rightarrow\vw\rightarrow\lab[\mu]$. For such a Markov chain, it is
easy to show the following data processing inequality \cite{s_cover2006},
\begin{equation}
  \Delta \mutual{\tlab[\mu]}{\vw} \ge \Delta \mutual{\tlab[\mu]}{\lab[\mu]},
\end{equation}
which completes our derivation.

\section{Hebbian learning in the thermodynamic limit}
\label{sec:hebbian-calculations}

In this section, we provide additional analytical calculations for Hebbian
learning in the thermodynamic limit for long times $t\rightarrow\infty$.



\subsection{Direct integration of the full distribution $p(\tlabs, \vw, \labs)$ }

To compute the mutual information between the true and predicted label of a
given sample, $\mutual{\tlab[\mu]}{\lab[\mu]}$, we need the distribution
$p(\tlab[\mu], \lab[\mu])$ or, since both $\tlab[\mu]$ and $\lab[\mu]$ are
symmetric binary random variables, the probability that
$\tlab[\mu] = \lab[\mu]$. Our aim in this section is to obtain this probability
for Hebbian learning in the thermodynamic limit with $t\rightarrow\infty$ by
direct integration of the full distribution over the true labels, weights and
predicted labels for a given set of samples $\{\vsample[\mu]\}$, which will also
give additional motivation for introducing the stability $\Delta^\mu$ of a
sample.

We start with the full probability distribution
\begin{equation}
  p(\tlabs, \vw, \labs) = \left(\frac{1}{2}\right)^P \left(\prod_{n=1}^N
    \frac{e^{-(\wn-\lr \mathcal{F}_n)^2/2}}{\sqrt{2\pi}}\right)
  \left(\prod_{\mu=1}^P \frac{e^{\lab[\mu]\vw\cdot\vsample[\mu]/2\sqrt{N}}}{e^{-\vw\cdot\vsample[\mu]/2\sqrt{N}}+e^{\vw\cdot\vsample[\mu]/2\sqrt{N}}}\right),
\end{equation}
where $\lr$ is the learning rate and $\mathcal{F}_n$ is a suitably scaled
average over the samples and labels,
\begin{equation}
  \mathcal{F}_n=\frac{1}{\sqrt{N}}\sum_{\rho=1}^P \tlab[\rho]\sample[\rho]_n
\end{equation}
While the sum over the predicted labels $\lab[\rho\neq\mu]=\pm1$ is trivial, we
can integrate over the true labels by noting that we can rewrite the exponent as
\begin{equation}
  p(\tlabs, \vw, \lab[\mu]) = \left(\frac{1}{2}\right)^P \left(\prod_{n=1}^N
    \frac{e^{-(\wn-\lr \tlab[\mu] \sample[\mu]_n/\sqrt{N}- \lr \mathcal{F}^{\overline{\mu}}_n)^2/2}}{\sqrt{2\pi}}\right)
   \frac{e^{\lab[\mu]\vw\cdot\vsample[\mu]/2\sqrt{N}}}{e^{-\vw\cdot\vsample[\mu]/2\sqrt{N}}+e^{\vw\cdot\vsample[\mu]/2\sqrt{N}}}
\end{equation}
where the only dependence of the weight distribution on the true labels
$\tlab[\rho\neq\mu]$ is now confined to the sum
\begin{equation}
  \mathcal{F}^{\overline{\mu}}_n\equiv\frac{1}{\sqrt{N}}\sum_{\rho\neq\mu}^P \tlab[\rho]\sample[\rho]_n.
\end{equation}
In the thermodynamic limit, this allows us to replace the sum over all
$\tlab[\mu\neq\rho]$ by an integral over the stochastic variable
$\mathcal{F}^{\overline{\mu}}_n$, which is normally distributed by the central
limit theorem and has mean 0 and variance $\alpha$. Carrying out the integral,
we find
\begin{equation}
  \label{eq:3}
  p(\tlab[\mu], \vw, \lab[\mu]) =  \left(\prod_{n=1}^N
    \frac{e^{-(\wn-\lr \tlab[\mu] \sample[\mu]_n/\sqrt{N})^2/2(1+\alpha\lr^2)}}{\sqrt{2\pi(1+\alpha\lr^2)}}\right)
  \frac{e^{\lab[\mu]\vw\cdot\vsample[\mu]/2\sqrt{N}}}{e^{-\vw\cdot\vsample[\mu]/2\sqrt{N}}+e^{\vw\cdot\vsample[\mu]/2\sqrt{N}}}
\end{equation}
Since both $\tlab[\mu]$ and $\lab[\mu]$ are binary random variables and
$\tlab[\mu]=\pm1$ with equal probabilities, the mutual information between the
true and predicted label can be written as
\begin{equation}
  \mutual{\tlab[\mu]}{\lab[\mu]} = \ln 2 - S[p(\tlab[\mu]=\lab[\mu])]
\end{equation}
with the shorthand for the binary entropy
$S[p]=-p \ln p - (1-p)\ln(1-p)$~\cite{s_cover2006}. With $\lab[\mu]=\tlab[\mu]$ in
the exponential term of eq.~\eqref{eq:3} and noting that
$(\tlab[\mu]\sample[\mu]_n)^2=1$ for all $\tlab[\mu]$, $\sample[\mu]_n$, we then
have
\begin{equation}
  \label{eq:4}
  p(\tlab[\mu] =\lab[\mu], \vw) =  \left(\prod_{n=1}^N
    \frac{e^{-(\wn \tlab[\mu] \sample[\mu]_n -\lr/\sqrt{N})^2/2(1+\alpha\lr^2)}}{\sqrt{2\pi(1+\alpha\lr^2)}}\right)
  \frac{e^{\tlab[\mu]\vw\cdot\vsample[\mu]/2\sqrt{N}}}{e^{-\vw\cdot\vsample[\mu]/2\sqrt{N}}+e^{\vw\cdot\vsample[\mu]/2\sqrt{N}}}
\end{equation}
It thus becomes clear that $\vw\cdot\vsample[\mu]\tlab[\mu]$ is the sum of $N$
random variables with mean $\lr/\sqrt{N}$ and variance $1+\alpha\lr^2$. We are
then motivated to introduce the stability of a sample,
\begin{equation}
  \Delta^\mu \equiv \frac{1}{\sqrt{N}} \vw\cdot\vsample[\mu]\tlab[\mu] = \act^\mu \tlab[\mu].
\end{equation}
which, from eq.~\eqref{eq:4}, is normally distributed with mean $\lr$ and
variance $1+\alpha\lr^2$. Introducing the stability allows us to replace the
integral over all the weights by an integral over the stability,
\begin{equation}
  p(\tlab[\mu] = \lab[\mu]) = \int_{-\infty}^\infty \dd \Delta^\mu
  \frac{e^{-(\Delta^\mu -\lr)^2/2(1+\alpha\lr^2)}}{\sqrt{2\pi(1+\alpha\lr^2)}}
  \frac{e^{\Delta^\mu}}{1+e^{\Delta^\mu}} = \int_{-\infty}^\infty \dd \Delta^\mu
  p(\Delta^\mu) \frac{e^{\Delta^\mu}}{1+e^{\Delta^\mu}}
\end{equation}
which is the distribution obtained as eq.~\eqref{eq:pR} of the main text.

\subsection{Direct derivation of the distribution of stabilities}

\label{sec:stabilities}

Let us quickly show how the distribution of stabilities
\begin{equation}
  \label{eq:supp_stability}
  \Delta^\mu \equiv \frac{1}{\sqrt{N}} \vw\cdot\vsample[\mu]\tlab[\mu],
\end{equation}
$\mu=1,\dots,P$, is obtained directly from its definition. The weights are given
by
\begin{equation}
  \label{eq:supp_mean-weight}
  \vw = \frac{1}{\sqrt{N}}\lr \sum_{\rho=1}^P \vsample[\rho]\tlab[\rho] + \mathbf{y}
\end{equation}
with $\mathbf{y}=(y_1,y_2,\dots,y_N)$ where $y_n$ are normally distributed
random variables with mean 0 and variance 1 arising from the thermal
fluctuations in equilibrium. Substituting eq. \eqref{eq:supp_mean-weight} into
\eqref{eq:supp_stability}, we have
\begin{align}
  \Delta^\mu = & \frac{1}{N} \lr \sum_{\rho=1}^P \tlab[\rho]\tlab[\mu]
                 \vsample[\rho]\cdot\vsample[\mu] +
                 \frac{1}{\sqrt{N}}\tlab[\mu]\vsample[\mu]\cdot\mathbf{y} \\
  = & \lr + \frac{1}{N}\lr\sum_{\rho\neq\mu}^P\tlab[\rho]\tlab[\mu]
      \vsample[\rho]\cdot\vsample[\mu] + \frac{1}{\sqrt{N}}\tlab[\mu]\vsample[\mu]\cdot\mathbf{y}
\end{align}
where going to the last line we have used the fact that
$\vsample[\mu]\cdot\vsample[\mu]=N$. By inspection, we see that the second term
is the sum of $N(P-1)\approx NP$ random numbers $\pm \lr / N$ and the last term is
the sum of $N$ random numbers $y_n/\sqrt{N}$. By the central limit theorem,
$\Delta^\mu$ is hence normally distributed with mean $\overline{\avg{\Delta^\mu}}=\lr$ and
variance
\begin{equation}
  \overline{\avg{(\Delta^\mu)^2}}-\overline{\avg{\Delta^\mu}}^2 = \lr^2 + NP \frac{\lr^2}{N^2} +
  N\frac{1}{N}-\lr^2 = 1+\alpha\lr^2.
\end{equation}

\subsection{Analytical approximation for $\mutual{\tlab}{\lab}$}

\label{sec:approximation}

We quantify the success of learning using the mutual information per sample,
\begin{equation}
  \mutual{\tlab[\mu]}{\lab[\mu]} = \ln 2 - S(p^\mu_\textrm{C})
\end{equation}
where $S(p)=-[p \ln p + (1-p)\ln(1-p)]$ is the binary Shannon entropy and
$p^\mu_\textrm{C}$ is defined as
\begin{equation}
  \label{eq:pC}
  p^\mu_\textrm{C}\equiv p(\lab[\mu]=\tlab[\mu]) = \int_{-\infty}^\infty \dd \! \Delta^\mu \; p(\Delta^\mu)\frac{e^{\Delta^\mu}}{e^{\Delta^\mu}+1}
\end{equation}
The stabilities $\Delta^\mu$ are normally distributed with mean $\lr$ and variance
$1+\alpha \lr^2$ (see section \ref{sec:stabilities}). This integral does not
have a closed-form analytical solution, but here we will demonstrate a very good
analytical approximation.

To that end, we first rewrite the sigmoid function in the integrand in terms of
the hyperbolic tangent and exploit the similarity of the latter to the error
function:
\begin{align}
  \pC^\mu = & \int_{-\infty}^\infty \dd \Delta^\mu\; p(\Delta^\mu)
          \frac{e^{\Delta^\mu/2}}{e^{\Delta^\mu/2}+e^{-\Delta^\mu/2}} \\
      = & \frac{1}{2} + \frac{1}{2}\int_{-\infty}^\infty \dd \Delta^\mu\; p(\Delta^\mu)
          \tanh(\Delta^\mu / 2) \\
    \simeq & \frac{1}{2} + \frac{1}{2}\int_{-\infty}^\infty \dd \Delta^\mu\; p(\Delta^\mu)
          \erf(\gamma \Delta^\mu / 2)
\end{align}
where we choose $\gamma=4/5$ by inspection of the graphs of the two
functions. Now the convolution of a normal distribution and an error function
has an exact solution,
\begin{equation}
  \frac{1}{\sqrt{2\pi d^2}}\int_{-\infty}^\infty \dd x \erf(ax+b)
  \exp\left(-\frac{(x-c)^2}{2d^2}\right)
  = \erf\left(\frac{b+ac}{\sqrt{1+2a^2d^2}}\right).
\end{equation}
Setting $a=\gamma/2$, $b=0$, $c=\lr$ and $d^2=1+\alpha\lr^2$, we find that
\begin{align}
  \pC^\mu(\alpha, \lr) \simeq & \frac{1}{2} + \frac{1}{2}\erf\frac{\gamma\lr/2}{\sqrt{1+\gamma^2(1+\alpha
                                \lr^2)/2}}\\
  = & \frac{1}{2} + \frac{1}{2}\erf\frac{\lr/2}{\sqrt{25/16+1/2+\alpha
      \lr^2/2}}\\   
  \simeq &  \frac{1}{2} + \frac{1}{2}\erf\frac{\lr/2}{\sqrt{2(1+\alpha
           \lr^2/4)}}\\
  = & p(\Delta^\mu > 0 | \alpha, \lr/2) \label{eq:approximation}
\end{align}
where in the last line we recognise by inspection that our result is nothing but
the integral over the distribution of stabilities $p(\Delta^\mu|\alpha, \lr/2)$ from
0 to $\infty$. The probability that the neuron predicts the correct label is
hence given by the probability that the neuron learned the label correctly,
$\Delta^\mu>0$, with \emph{half the learning rate}.

\bibliographystyle{aip}

\end{document}